\documentclass[10pt,aps,prd,twocolumn,
nofootinbib,noshowkeys,noshowpacs,
superscriptaddress,floatfix]{revtex4-2}
\usepackage[utf8]{inputenc}
\usepackage[T1]{fontenc}
\usepackage{amsfonts,amsthm,tensor}
\usepackage{amsmath}
\usepackage{amssymb}
\usepackage[dvips]{graphics,graphicx}
\usepackage[usenames,dvipsnames]{color}
\definecolor{darkblue}{RGB}{0,0,196}
\definecolor{darkgreen}{RGB}{0,120,0}
\usepackage[colorlinks=true,linktocpage=true,linkcolor=darkblue,citecolor=red,urlcolor=darkblue]{hyperref}
\usepackage{cancel}
\usepackage{bbold}
\usepackage{subfigure}
\usepackage{stix}
\usepackage{multirow}
\usepackage{longtable}
\usepackage{color}
\usepackage[normalem]{ulem}
\usepackage{bigints}
\usepackage{xparse}
\usepackage{verbatim}
\usepackage{minibox}
\usepackage{comment}
\usepackage{appendix}
\usepackage{scalerel}
\usepackage{marginnote}
\usepackage{graphicx}
\usepackage[nice]{nicefrac}
\usepackage{soul}

\usepackage{hepunits}
\begin{document}
\preprint{}

\title{Study of exclusive decays of \texorpdfstring{$B_s \to \psi(1S,2S) K_s$}{Lg} and \texorpdfstring{$B_s \to \eta_c(1S,2S) K_s$}{Lg}}
\author{Lopamudra Nayak}
\email{lopalmn95@gmail.com}
\affiliation{National Institute of Science Education and Research, An OCC of Homi Bhabha National Institute, Bhubaneswar, Odisha, India}

\author{Sonali Patnaik}
\email{sonali\_patnaik@niser.ac.in}
\affiliation{National Institute of Science Education and Research, An OCC of Homi Bhabha National Institute, Bhubaneswar, Odisha, India}

\author{Priyanka Sadangi}
\email{priyanka.sadangi@niser.sc.in}
\affiliation{National Institute of Science Education and Research, An OCC of Homi Bhabha National Institute, Bhubaneswar, Odisha, India}

\author{Sanjay Kumar Swain}
\email{sanjay@niser.ac.in}
\affiliation{National Institute of Science Education and Research, An OCC of Homi Bhabha National Institute, Bhubaneswar, Odisha, India}
\begin{abstract}
We analyze the exclusive two-body nonleptonic decays of $B_s$ meson to ground as well as radially excited $2S$ charmonium state with a light meson $K_s$, induced by the $b\to c\bar{c}d$ transition. Within the framework of relativistic independent quark (RIQ) model based on a flavor-independent interaction potential in scalar-vector harmonic form, we calculate the weak form factors from the overlapping integrals of meson wave function obtained in this model. Using the factorization approximation, we predict the branching fractions for the $B_s \to \psi(1S,2S) K_s$ and $B_s \to \eta_c(1S,2S) K_s$, which can be compared with future theoretical predictions. Branching fraction for $B_s\to J/\psi K_s$ decay is found to be in good agreement with the data from LHCb Collaboration, whereas for $B_s\to \psi(2S) K_s$, it is found to be within the detection ability of the CMS Collaboration. We also predict the ratios of branching fractions $(\cal{R})$, which are in broad agreement with the data from LHCb Collaboration. These results indicate that the present approach works well in the description of exclusive nonleptonic $B_s$ decays within the framework of the RIQ model.  	
\end{abstract}
\maketitle
\section{Introduction}
\label{sec:intro}

The exclusive nonleptonic decays of the heavy meson $B_s$ have provided a precise and consistent picture of the flavor sector of the Standard Model (SM) over the past decade \cite{BaBar:2014omp}. Several of these decay channels of $B_s$ meson into charmonium resonance have increased the interest, as it is well suited to study the flavor sector of the SM and to search for indication of new physics (NP),  beyond the SM (BSM). Among the possibilities of nonleptonic $B_s$ decay modes, the Cabibbo–Kobayashi–Maskawa (CKM)-suppressed mode induced by quark level transition $b\to c\bar{c}d$ that involve a charmonium meson and a neutral kaon meson in final state are of particular interest. Specifically, the charmonium vector meson $J/\psi$ is valuable in experimental studies due to its clean signal reconstruction $J/\psi \to \mu^+ \mu^-$ \cite{BaBar:2014omp}. Additionally, kaons are of significant experimental interest as they can be identified through their decay into a $\pi^+\pi^-$ pair coming from a displaced vertex. The decays of neutral $B_s$ meson into a charmonium state and a kaon have received substantial attention because these channels provide techniques for investigating CP violation and determining the CKM matrix angles within the unitarity triangle. They also serve as avenues to test various Quantum Chromodynamics (QCD) -inspired models and explore potential deviations from the physics BSM. Furthermore, the $B_s\to J/\psi K_s$ decay mode is used as control channel~\cite{DeBruyn:2014oga} to handle the phase shift $\Delta\varphi_d$~\cite{Fleischer:2012dy} which is the theoretical uncertainty of  the $B_d^0-\bar{B_d}^0$ mixing phase $\varphi_d$, arising from doubly Cabibbo-suppressed penguin contributions. In the SM, $\varphi_d$ is same as $2\beta$. For measuring the CKM angle $\sin{2\beta}$, the decay $B_d \to J/\psi K_s$ is considered as the "golden mode". This decay is related to the $B_s\to J/\psi K_s$ decay through the swapping of all $d$ and $s$ quarks, a relationship governed by U-spin symmetry, a subgroup of SU(3) \cite{Fleischer:1999nz}. This symmetry connects the strong interaction dynamics of both decay modes, allowing for a comparative analysis of the penguin contributions. By studying these decays, one can accurately quantify and correct for the penguin-induced phase shift, thereby refining the measurement of $\varphi_d$. \\

Experimentally, the first observation \cite{CDF:2011fhd} of the Cabibbo suppressed decay $B_s \to J/\psi K_s$ has been reported by CDF Collaboration, using a sample derived from an integrated luminosity of 5.9 fb$^{-1}$ of proton-antiproton collisions at center-of-mass energy, $\sqrt{s}=1.96$ TeV produced at the Fermilab's Tevatron in 2011. In 2012, a more precise measurement of this branching fraction (BF) is presented by LHCb experiment \cite{LHCb:2012irf} using integrated luminosity of 0.41 fb$^{-1}$ of proton-proton ($pp$) collision at $\sqrt{s}=7$ TeV, and observed the ratio of BFs,
\vspace{-2mm}
\begin{align*}
\frac{{\cal B}(B_s\to J/\psi K_s)}{{\cal B}(B_d\to J/\psi K_s)} = 0.042 &\pm 0.0049\; (stat)\pm 0.0023\;(syst)\\\nonumber
&\pm 0.0033\; \big(\frac{f_s}{f_d}\big), 
\end{align*}
\vspace{-5mm}

\noindent where $\frac{f_s}{f_d}$ is ratio of fragmentation fractions of the $B_s$ and $B_d$. The above mentioned uncertainties include statistical, systematic, and those arising from the uncertainty in $f_s/f_d$, respectively. Using the world average BF of $B_d\to J/\psi K_s$, LHCb gets the time integrated BF of $B_s \to J/\psi K_s$ as:
\vspace{-3mm}
\begin{align*}
{\cal B}(B_s \to J/\psi K_s) = 
[1.83 &\pm 0.21(stat)\pm 0.10(syst)\pm 0.14\;(\frac{f_s}{f_d})\\\nonumber
&\pm 0.07\;(B_d\to J/\psi K_s)]\times 10^{-5}.
\end{align*}
An updated measurement of the ratio of BFs,
\vspace{-1mm}
\begin{align*}
\frac{{\cal B}(B_s\to J/\psi K_s)}{{\cal B}(B_d\to J/\psi K_s)} = 0.0431 &\pm 0.0017\; (stat)\pm 0.0012\;(syst)\\\nonumber
&\pm 0.0025\; (\frac{f_s}{f_d}), 
\end{align*}
\vspace{-4mm}

\noindent is presented by LHCb Collaboration in 2015 \cite{LHCb:2015brj}, by using which they measured the BF of $B_s\to J/\psi K_s$ as $[1.93 \pm 0.08(stat)\pm 0.05(syst)\pm 0.11\big(\frac{f_s}{f_d}\big)
\pm 0.07\;(B_d\to J/\psi K_s)]\times10^{-5}$. The results are obtained using data corresponding to an integrated luminosity of 3.0 fb$^{-1}$ of $pp$ collisions recorded with the LHCb detector at $\sqrt{s}$ = 7 and 8 TeV.\\

Moreover, recently, the first measurement of $B_s\to \psi(2S)K_s$ decay, using a data sample of $pp$ collisions at $\sqrt{s}$ = 13 TeV in 2017 and 2018 with an integrated luminosity of 103 fb$^{-1}$ has been published \cite{CMS:2022cot} in 2022. They measured the ratio of BF of $B_s\to \psi(2S) K_s$ decay to the $B_d\to \psi(2S) K_s$ as: 

\vspace{-5mm}

\begin{align*}
\frac{{\cal B}(B_s\to \psi(2S) K_s)}{{\cal B}(B_d\to \psi(2S) K_s)} = \big[3.33 &\pm 0.69\; (stat)\pm 0.11\;(syst)\\\nonumber
&\pm 0.22\; (\frac{f_s}{f_d})\big]\times 10^{-2} ,
\end{align*}
where the last uncertainty is related to the used value $\frac{f_s}{f_d}=0.208\pm0.021$. Using the world average value of ${\cal B}(B_d\to \psi(2S) K_s)=(2.90\pm0.25)\times 10^{-4}$ \cite{ParticleDataGroup:2020ssz}, the BF of $B_s\to \psi(2S)K_s$ has been evaluated as $\big[0.97\pm0.20 (stat)\pm 0.03 (syst)\pm 0.22\big(\frac{f_s}{f_d}\big)\pm 0.08({B_d\to \psi(2S) K_s})]\times 10^{-5}$.\\

In contrast to $B_s\to J/\psi K_s$, the $\eta_c$ must be reconstructed from the hadronic decays rather than from a leptonic final state with relatively low combinatorial background.
The decay $B_s\to \eta_c K_s$ proceed through the same $b\to c\bar{c}d$ quark diagram as the mode $B_s\to J/\psi K_s$, used to measure the CP violating parameter with reasonable theoretical uncertainty. Upto now, the experimental information on $B_s$-decays into $\eta_c$ has been sparse. With the recent progress in the measurement \cite{CMS:2022cot} of nonleptonic $B_s\to \psi(2S)K_s$ decay channel, the exclusive decay $B_s\to \eta_cK_s$ could be within experimental reach very soon.\\

Describing nonleptonic decay is theoretically challenging due to the strong influence of confining color forces and the involvement of matrix elements of local four-quark operators within the non-perturbative QCD approach, a mechanism still not thoroughly understood within the SM framework. When excluding weak annihilation contributions, the nonleptonic transition amplitudes are conveniently approximated using the so-called naive factorization method \cite{Barik:2001vp,Barik:2009zzb,Bauer:1986bm,Beneke:1999br,Beneke:2000ry,Beneke:2001ev,Beneke:2003zv,Thomas:2005bu}. This approach works reasonably well in two-body nonleptonic $B_s$ decays, particularly where the quark-gluon sea is suppressed within heavy quarkonium \cite{Deandrea:1993ma,Albertus:2014mra,Colangelo:2010wg,Parkhari:2023fne,Morales:2016pcq}. Bjorken's original argument on color transparency \cite{Bjorken:1988kk}, theoretical advancements based on QCD in the limit of large number of colors ($\frac{1}{N_c}$) \cite{Buras:1985xv}, and the Heavy Quark Effective Theory (HQET) \cite{Neubert:1993mb} provide rationale for this approximation. In this study, our focus lies solely on the impact of current-current operators \cite{Neubert:1997uc} in calculating the dominant tree-level diagram for these decays. While the penguin diagram's influence may be notable in assessing CP-violation and probing for BSM physics, its significance to the decay amplitudes under consideration here is deemed less pronounced.\\

In our current analysis, we adopt a phenomenological model framework known as the relativistic independent quark (RIQ) model, which has been applied in wide-ranging hadronic sector describing the static properties of hadrons \cite{Barik:1986mq,Barik:1987zb,Barik:1993aw} and their decay properties in the radiative, weak radiative, rare radiative \cite{Barik:1992pq,Barik:1994vd,Priyadarsini:2016tiu,Barik:1995sq,Barik:2001gr,Barik:1996kn}; leptonic, weak leptonic, radiative leptonic \cite{Barik:1996kn,Barik:1993yj,Barik:1993aw,Barik:2008zza,Barik:2008zz} and semileptonic \cite{Barik:2009zza,Barik:1996xf,Barik:1997qq} decays of mesons. We have predicted the magnetic dipole and electromagnetic transitions of $B_c$ and $B_c^*$ mesons in their ground as well as excited states\cite{Patnaik:2017cbl,Patnaik:2018sym}; the exclusive semileptonic $B_c$-mesons decays to the charmonium ground states in the vanishing \cite{Patnaik:2019jho} and non-vanishing \cite{Nayak:2021djn,Nayak:2022gdo,Patnaik:2023efe,Patnaik:2023ins} lepton mass limit. In this model our group have predicted \cite{Barik:2001vp,Barik:2009zzb,Naimuddin:2012dy,Kar:2013fna,Nayak:2022qaq,Dash:2023hjr} the exclusive two body nonleptonic decays of heavy flavored meson to the charmonium and charm mesons in their ground as well as radially excited states. This has greatly inspired us to broaden the utilization of the RIQ model to examine the two-body nonleptonic decay of $B_s\to \psi(nS) K_s$ and $B_s\to \eta_c(nS) K_s$, where $n=$ 1 or 2.\\

The rest of the paper is organized as follows. In section \ref{H_eff_sec}, we present the effective Hamiltonian and factorization for two-body nonleptonic decay modes of $B_s$ mesons induced by $b\to c\bar{c}d$ transitions at the quark level. Section \ref{model_expressions} elaborates on the weak decay form factors, which represent the hadronic amplitudes, and are computed from the overlap integral of meson wave-functions within the framework of the RIQ model. Section \ref{Num_analysis} is dedicated to the presentation of numerical results and subsequent discussion, while Section \ref{sum_con} provides a concise summary and conclusion. Additionally, the Appendix offers a brief overview of the RIQ model framework, the wave packet representation of meson states, and the momentum probability amplitudes of constituent quarks within the meson bound-state. 

\section{Effective Hamiltonian and factorization approximation}
\label{H_eff_sec}
The effective weak Hamiltonian for two body nonleptonic $B_s$ decays, induced by $b\to c\Bar{c}d$ decay at the quark level, consists of a sum of local four quark operators ${\cal O}_i$ multiplied by short distance coefficients ${\cal C}_i$ and products of elements of the quark mixing matrix $\lambda_c=V_{cb}V_{cd}^*$, can be written as: 
\begin{eqnarray}
    {\cal H}_{eff}=&& \frac{G_F}{\sqrt{2}} \lambda_c\sum {\cal C}_i {\cal O}_i\nonumber\\
    =&& \frac{G_F}{\sqrt{2}}V_{cb}V_{cd}^* \big[{\cal C}_1(\mu)(\bar{c}b)(\bar{d}c)+{\cal C}_2(\mu)(\bar{d}b)(\bar{c}c)\big]+H.c.\nonumber\\
    \label{equn:Hameffect}
\end{eqnarray}
Here, $G_F$ is the Fermi coupling constant, $V_{ij}$ are CKM factors,  $(\bar{q}_{\alpha}q_{\beta})$ is a short notation for $V-A$ current $q_{\alpha}\gamma^{\mu}(1-\gamma_5)q_{\beta}$, and ${\cal C}_{1,2}$ are the Wilson coefficients. As the decays: $B_s \to J/\psi (\eta_c)K_s$, in ground and radially excited (2S) charmonium states, are related by a tree-level amplitude proportional to $V_{cb}V_{cd}^*$, so we consider here only the tree-level current-current operators (${\cal O}_{1,2}$) by neglecting the penguin diagrams.\\ 

In the framework of naive factorization, the nonleptonic decay amplitude is approximated by the product of two matrix elements of quark currents as:
  \begin{equation}
  \label{eq:amplitude}
  	\langle M_1M_2|{\cal O}|B_{s}^0\rangle_i=\langle M_1|J^{\mu}|0\rangle \langle M_2|J_{\mu}|B_{s}^0\rangle+(M_2\leftrightarrow M_1),
  \end{equation}
where $J_\mu$ is the weak current and $M_i$ is a pseudoscalar $(P)$ or a vector $(V)$ meson. One of these is the matrix element for the $B_s$ transition to one final mesons state, while the other matrix element corresponds to the transition from the vacuum to other final meson state. The latter is given by the corresponding meson decay constant. In this way the hadronic matrix element of four-quark operators can be expressed as the product of decay constant and invariant weak form factors \cite{Bediaga:2011cs,Hernandez:2006gt,PhysRevD.63.114002,PhysRevD.80.014004,PhysRevD.86.094028,Kar:2013fna,Beneke:1999br,Beneke:2001ev,Thomas:2005bu,Buchalla:2008jp,Buras:1998raa}.\\

Of course, there is difficulty inherent in such an approach because the Wilson's coefficients, which include the short distance QCD effect between $\mu=m_N$ and $\mu=m_b$ are $\mu$-scale and renormalization scheme dependent, while $\langle {\cal O}\rangle_i$ are $\mu$-scale and renormalization scheme independent. As a result, the physical amplitude depends on the $\mu$ scale. However, the naive factorization disentangles the long distance effects from the short distance sector assuming that the matrix element $\langle {\cal O}\rangle$ at $\mu$ scale, contain non-factorizable contributions in order to cancel the $\mu$ dependence and scheme dependence of $c_i(\mu)$ \cite{Beneke:1999br,Beneke:2000ry,Beneke:2001ev,Beneke:2003zv,Thomas:2005bu,Buchalla:2008jp,Buras:1998raa}. In general, it works in some two-body nonleptonic decays of heavy mesons in the limit of a large number colors. It is expected that the factorization scheme works reasonably well in two-body nonleptonic $B_s$ decays with charmonium meson in the final states \cite{Morales:2016pcq,Colangelo:2010wg}.\\
     
We also neglect here the so-called $W$-exchange and annihilation diagram, since in the limit   $M_W\to \infty$, they are connected by Fiertz transformation and are doubly suppressed by the kinematic factor of the order $(\frac{M_i^2}{M_W^2})$. We also discard the color octet current which emerge after the Fiertz transformation of color-singlet operators. Clearly, these currents violate factorization since they can not provide transitions to the vacuum states. Taking into account the Fiertz reordered contribution, the relevant coefficients are not $C_1(\mu)$ and $C_2(\mu)$ but the combination:
\vspace{-4mm}
     \begin{equation}
     	a_{1,2}(\mu)=C_{1,2}(\mu)+\frac{1}{N_c}C_{2,1}(\mu),
     \end{equation}
where $N_c$ is color factor. Assuming large $N_c$ limit to fix the QCD coefficients $a_1 \approx C_1$ and $a_2\approx C_2$ at $\mu\approx m_b^2$, nonleptonic decays of heavy mesons have been analyzed in \cite{Colangelo:1999zn,Neubert:1997uc}.\\
  
The matrix elements corresponding to the transition from vacuum to one of the final state pseudoscalar or vector meson are covariantly expanded in terms of the meson decay constant $f_{P,V}$ as:
\vspace{-5mm}
\begin{eqnarray}
\langle P|\bar{q}_i^{'}\gamma^\mu \gamma_5 q_j|0\rangle=if_Pp_P^\mu\nonumber\\
\langle V|\bar{q}^{'}_i\gamma^\mu q_j|0\rangle=e^{*\mu}f_Vm_V
\end{eqnarray}
The covariant decomposition of matrix elements of the weak current $J_\mu$ between initial and final pseudoscalar  meson state is: 
  \begin{eqnarray}
  \label{Lor_inv_ff_pseudo}
  	&&\langle P(p_P)|\bar{q}_{c}\gamma_\mu q_{b}|B_s(p)\rangle\nonumber\\&&=\big[(p+p_P)_\mu-\frac{M^2-m_P^2}{q^2}q_\mu\big]F_1(q^2)+\frac{M^2-m_P^2}{q^2}q_\mu F_0(q^2)\nonumber\\
  	&&=(p+p_P)_\mu f_+(q^2)+(p-p_P)_\mu f_-(q^2),
  \end{eqnarray}
where
\begin{eqnarray}
\label{fp}
	f_+(q^2)=&&F_1(q^2)\\
 \label{fm}
	f_-(q^2)=&&\frac{M^2-m^2_P}{q^2} \big[F_0(q^2)-F_1(q^2)\big]
\end{eqnarray}
 For transition to the vector meson final state, corresponding matrix element is parametrized as:
\begin{equation}
	\langle V(p_V)|\bar{q}_c\gamma_\mu q_b|B_s(p)\rangle=\frac{2V(q^2)}{M+m_V}i\epsilon_{\mu\nu\rho\sigma}\ e^{*\nu} p^\rho p_V^{\sigma}
\end{equation}
and
\begin{eqnarray}
	&&\langle V(p_V)|\bar{q}_c\gamma_\mu\gamma_5q_b|B_s(p)\rangle\nonumber\\&&=(M+m_V)e^*_\mu A_1(q^2)-\frac{A_2(q^2)}{M+m_V}(e^*.q)(p+p_V)_\mu\nonumber\\
	&&-2m_V\frac{e^*.q}{q^2}q_\mu A_3(q^2)+2m_V\frac{e^*.q}{q^2}q_\mu A_0(q^2),
\end{eqnarray}
where \begin{equation}
	A_3(q^2)=\frac{M+m_V}{2m_V}A_1(q^2)-\frac{M-m_V}{2m_V}A_2(q^2)
\end{equation}
Here, $p \text{and} p_{P,V}$ stand for the four momentum of the initial and final state meson, respectively. M is the mass of decaying $B_s$ and $m_P$ and $m_V$ stand for the mass of the pseudoscalar and vector mesons, respectively, in the final state. $q=p-p_{P,V}$ denotes the four momentum transfer and $\hat{e}^*$ is the polarization of the final state vector meson. In order to cancel the poles at $q^2=0$, invariant weak form factors: $F_0(q^2), F_1(q^2),A_0(q^2)$ and $A_3(q^2)$ satisfy following conditions: 
\begin{equation}
	F_0(0)=F_1(0)\ \ \ \ \text{and} \ \ \ A_0(0)=A_3(0).\nonumber
\end{equation}
For $B_s\to PV$ transition, the corresponding decay rate is expressed in terms of the decay amplitude $A(B_s\to PV)$ as:
\begin{equation}
	\Gamma(B_s\to PV)=\frac{|\vec{k}|^3}{8\pi m_V^2}|A(B_s\to PV)|^2,
\end{equation}
where $|\vec{k}|$ is the magnitude of three-momentum of the final state meson. In the parent meson rest frame it is given by 
\begin{equation}
|\vec{k}|=\Bigg[\Big(\frac{M^2+m_P^2-m_V^2}{2M}\Big)^2-m_P^2\Bigg]^{1/2}
\end{equation}
The decay rate for nonleptonic transition $B_s\to P_1P_2$ is expressed in terms of the decay amplitude $A(B_s\to P_1P_2)$ as:
\begin{equation}
	\Gamma(B_s\to P_1P_2)=\frac{|\vec{k}|}{8\pi M^2}|A(B_s\to P_1P_2)|^2,
\end{equation}
where $|\vec{k}|$ is the magnitude of three-momentum of the final state meson. In the parent meson rest frame it is given by 
\begin{equation}
|\vec{k}|=\Bigg[\Big(\frac{M^2+m_{P_1}^2-m_{P_2}^2}{2M}\Big)^2-m_{P_1}^2\Bigg]^{1/2}
\end{equation}

The relevant decay amplitude (\ref{eq:amplitude}) can either be expressed in the following form:
\begin{eqnarray}
A=&&\frac{G_F}{\sqrt{2}}(\text{CKM factor})(\text{QCD factor})\nonumber\\
&&\times\, (M^2-m_{P_{1,2}}^2)f_{P_{2,1}}F_0^{B_s\to P_{1,2}}(q^2),\nonumber\\
\text{and}\nonumber\\	A=&&\frac{G_F}{\sqrt{2}}(\text{CKM factor})(\text{QCD factor})\nonumber\\
&&\times \, 2m_Vf_VF_1^{B_s\to P}(q^2)\nonumber
\end{eqnarray}

\noindent for $B_s\to P_1P_2$ and $B_s\to PV$ decay, respectively. Here, the factorized amplitudes (\ref{eq:amplitude}) are expressed in terms of meson decay constants $(f_{P,V})$ and weak form factors $ F_0 \text{ and } F_1$. It is then straight forward to predict the decay rate for different decay processes in the RIQ model framework.

\vspace{-4mm}

\section{transition amplitude and weak decay form factors in the relativistic independent quark model}
\label{model_expressions}
\begin{figure}[hbt!]
\centering
\includegraphics[width=0.5\textwidth]{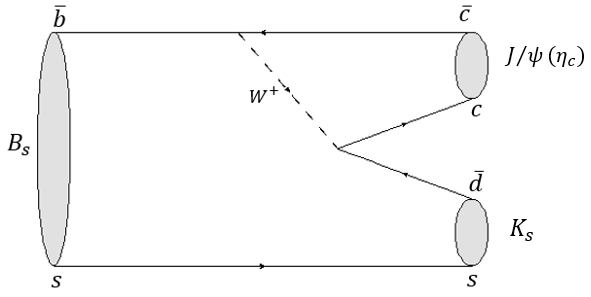}
\caption{Feynman Diagram contributing to the $B_s\to J/\psi(\eta_c) K_s$ Class II type decay.}
\label{fig:enter-label2}
 \end{figure}

We study two-body nonleptonic $B_s$ decays of categories: $B_s\to P_1P_2$ and $B_s\to PV$. The decay amplitude is calculated here from relevant tree-level "class-II" type diagram as shown in Fig. \ref{fig:enter-label2}. The "class-II" decay is characterized by internal $W$-emission, where the decay amplitude is proportional to the QCD-factor $a_2(\mu)$. As described above we consider the two-body nonleptonic $B_s$ decays induced by $b\to c\bar{c}d$ transition at quark level, with $s$-quark remaining a spectator. In the present study, we restrict our discussion to class II, $B_s\to \psi (nS) K_s$ and $B_s\to \eta_c (nS) K_s$ decay modes.
 
In fact the decay process physically occurs in the momentum eigenstates of participating mesons. Therefore, in a field-theoretic description of the decay process, it is appropriate to represent the meson bound state in terms of a momentum wavepacket reflecting momentum and spin distribution between constituent quark and antiquark in the meson core. In RIQ model, the wave packet corresponding to a meson bound state $|B_s(\vec{p},S_{B_s})\rangle$, for example, at a definite momentum $\vec{p}$ and spin state $S_{B_s}$ is represented by:
 \begin{eqnarray}
 \label{meson_state}
 |B_s(\vec{p},S_{B_s})\rangle =&&\;\hat{\Lambda}(\vec{p},S_{B_s})|(\vec{p}_{\bar{b}},\lambda_{\bar{b}});(\vec{p}_s,\lambda_s)\rangle\\
 	=&&\;\hat{\tilde{b}}^{\dagger}_b(\vec{p}_b,\lambda_b)\hat{b}_s^{\dagger}(\vec{p}_s,\lambda_s)|0\rangle, \nonumber
 \end{eqnarray}
where $|(\vec{p}_{\bar{b}},\lambda_{\bar{b}});(\vec{p}_s,\lambda_s)\rangle$ is the Fock-space representation of the unbound quark and antiquark in a color-singlet configuration with respective momentum and spin: $(\vec{p}_b,\lambda_b)$ and $(\vec{p}_s\lambda_s)$. $\hat{b}^{\dagger}_q(p_b,\lambda_b)$ and $\hat{\tilde{b}}_s(\vec{p}_s,\lambda_s) $ are the quark and antiquark creation operator. Here, $\hat{\Lambda}(\vec{p},S_{B_s})$ is a bag like integral operator taken in the form:

\vspace{-5mm}

\begin{eqnarray}
\label{bag_operator}
	\hat{\Lambda}(\vec{p},S_{B_s})=&&\frac{\sqrt{3}}{\sqrt{N(\vec{p})}}\sum_{\lambda_b,\lambda_s}\zeta^{B_s}_{b,s}(\lambda_b,\lambda_s)\int d^3p_bd^3p_s\nonumber\\
 && \times\; \delta^{(3)}(\vec{p}_b+\vec{p}_s-\vec{p}){\cal G}_{B_s}(\vec{p}_b,\vec{p}_s),
\end{eqnarray}
where $\sqrt{3}$ is the effective color factor and $\zeta^{B_s}_{b,s}(\lambda_b,\lambda_s)$ is the SU(6) spin-flavor coefficients for $B_s$ meson state. Imposing the normalization condition in the form $\langle B_s(\vec{p})|B_s(\vec{p}^{'}) \rangle =\delta^{(3)}(\vec{p}-\vec{p}^{'})$, the meson state normalization $N(\vec{p})$ is obtainable in an integral form:
\begin{equation}
	N(\vec{p})=\int d^3\vec{p}_b |{\cal G}_{B_s}(\vec{p}_b,\vec{p}-\vec{p}_b)|^2
\end{equation}
 ${\cal G}_{B_s}(\vec{p}_b,\vec{p}-\vec{p}_b)$ denotes the momentum distribution function for the quark and antiquark pair in the meson core. In this model ${\cal G}_{B_s}(\vec{p}_b,\vec{p}-\vec{p}_b)$ is taken in the form: ${\cal G}_{B_s}(\vec{p}_b,\vec{p}-\vec{p}_b)=\sqrt{G_b(\vec{p}_b)G_s(\vec{p}-\vec{p}_b)}$ in the straight forward extension of the ansatz of Margolis and Mendel in their bag model description \cite{Margolis:1983wz}; where $G_b(\vec{p}_b)$ and $G_s(\vec{p}-\vec{p}_b)$ refer to individual momentum probability amplitude of the constituent quark. Here, the effective momentum distribution function in fact embodies the bound state character in $|B_s(\vec{p},S_{B_s})\rangle$.

Any residual internal dynamics responsible for the decay process can, therefore, be described at the constituent level by the unbound quark and antiquark using usual Feynman technique. The constituent level $S$-matrix element $S_{fi}^b\to c\bar{c}d$ obtained from the appropriate Feynman diagram, when operated upon by the bag like operator $\hat{\Lambda}(\vec{p},S_{B_s})$ (\ref{bag_operator}) can give rise to the mesonic level $S$-matrix  in the form:
\begin{equation}
	S_{fi}^{B_s\to J/\psi(\eta_c) K_s}\to \hat{\Lambda}(\vec{p},S_{B_s})S_{fi}^{b\to c \bar{c}d}
\end{equation}
The decay modes $B_s\to J/\psi(\eta_c) K_s$  are induced by $b\to c\bar{c}d$ transition at quark level with emission of $W$-boson. The internally emitted $W$-boson with four momentum $q$, which decays to a quark-antiquark pair, subsequently the quark hadronize with the resulting anti-quark $c$ from $b$-antiquark to form the $S$-wave charmonium states. While the antiquark hadronize with the spectator quark $s$ to form the $K_s$. Considering the wave packet representation of the participating meson states in the factorized decay amplitude (\ref{eq:amplitude}), the  $S$-matrix element for $B_s\to J/\psi(\eta_c) K_s$ can be obtained in the general form:   
\begin{equation}
\label{S_fi}
	S_{fi}=(2\pi)^4 \delta^{(4)}(p-q-k)(-i{\cal M}_{fi})\times\frac{1}{\sqrt{V2E_{B_s}}}\prod_f \frac{1}{V2E_f}
\end{equation}  

\noindent The invariant transition amplitude ${\cal M}_{fi}$ is infact extracted in the form:

\vspace{-7mm}

\begin{equation}
	{\cal M}_{fi}=\frac{G_F}{\sqrt{2}}V_{bc}V_{cd}a_2 A,
\end{equation}
where $A=h^{\mu}H_{\mu}$with
\noindent\begin{equation}
h^{\mu}=\sqrt{\frac{2E_{J/\psi(\eta_c)}}{(2\pi)^3}}\langle J/\psi(\eta_c)(\vec{p}_2,S_{p_2})|J^{\mu}|0\rangle
\end{equation}
and 
\begin{eqnarray}
\label{H_mu}
	H_{\mu}=&&\frac{1}{\sqrt{N_{B_s}(\vec{p})N_{K_s}(\vec{k})}}\nonumber\\
 &&\int{\frac{d^3{\vec{p}_b}{\cal G}_{B_s}(\vec{p}_b,\vec{p}-\vec{p}_b){\cal G}_{K_s}(\vec{p}_b+\vec{k}-\vec{p},\vec{p}-\vec{p}_b)}{\sqrt{E_b(\vec{p}_b)E_d(\vec{p}_b+\vec{k}-\vec{p})}}}\nonumber\\
	&&\times\sqrt{\big[E_b(\vec{p}_b)+E_d(\vec{p}-\vec{p}_b)\big]\big[E_d(\vec{p}_b+\vec{k}-\vec{p})+E_{\bar{s}}(\vec{p}-\vec{p}_b)\big]}\nonumber\\
 &&\langle S_{K_s}|J_{\mu}(0)|S_{B_s}\rangle
\end{eqnarray}
Here, $E_b(\vec{p}_b)$ and $E_d(\vec{p}_b+\vec{k}-\vec{p})$ denote the energy of the non-spectator quark in the parent and daughter meson, respectively. $\vec{p}_b$, $\vec{p}$, $\vec{k}(\vec{p_1})$, and $\vec{p_2}$ represent the momentum vectors of the non-spectator constituent quark $b$, the parent meson $B_s$, the daughter meson $K_s$, and the charmonium state $J/\psi(\eta_c)$ respectively. $q$ stands for the four-momentum transfer associated with the process. Symbolically, $\langle S_{K_s}|J_{\mu}|S_{B_s}\rangle$ represents the spin matrix elements of the effective vector-axial vector current.

It is noteworthy that in the description of decay processes like $B_s\to J/\psi(\eta_c) K_s$ within the RIQ model framework, three-momentum conservation is explicitly ensured through $\delta^{(3)}(\vec{p}_{q_1}+\vec{p}_{q_2}-\vec{p})$ in the involved meson states. However, energy conservation in such a scheme is not explicitly ensured. This is a common issue in potential model descriptions of mesons as bound states of valence quarks and antiquarks interacting via some instantaneous potential. This issue has been addressed in our prior analyses concerning radiative leptonic decays of heavy-flavored mesons: $B$, $B_c$, $D$, and $D_s$ \cite{Barik:1993yj,Barik:1993aw,Barik:2008zza,Barik:2008zz}. In those analyses, the effective momentum distribution function ${\cal G}_M(\vec{p}_{q_1},\vec{p}_{q_2})$ encapsulating the bound-state characteristics of the meson ensures energy conservation in an average sense, satisfying $E_M=\langle M(\vec{p},S_M)|\big[E_{q_1}(\vec{p}_{q_1})+E_{q_2}(\vec{p}_{q_2})\big]|M(\vec{p},S_M)\rangle$. Therefore, we impose the energy conservation constraint: $M=E_{q_1}(\vec{p}_{q_1})+E_{q_2}(-\vec{p}_{q_1})$, where $M$ denotes the mass of the meson at rest. Along with the three-momentum conservation via appropriate $\delta^{(3)}(\vec{p}_{q_1}+\vec{p}_{q_2}-\vec{p})$ in the meson state, this ensures the required four-momentum conservation: $\delta^{(4)}(p-k-q)$ at the mesonic level, which is separated from the quark level integration to obtain the $S$-matrix element in the standard form (\ref{S_fi}).

Since the axial vector current does not contribute to the decay amplitude in the $B_s\to J/\psi(\eta_c) K_s$ decay processes, the only non-vanishing vector current part of (\ref{H_mu}) is simplified after calculating corresponding spin matrix elements using usual spin algebra. The resulting time-like and space-like part of the hadronic matrix element in the parent meson rest-frame are obtained, respectively, as:
 \begin{eqnarray}
 \label{V0}
 	\langle K_s(\vec{k})|V_0|B_s(0)\rangle =H_0=&&\int d\vec{p}_bC(\vec{p}_b)\bigl\{\big[E_b(\vec{p}_b)+m_b\big]\nonumber\\
  &&\big[E_d(\vec{p}_b+\vec{k})+m_d\big]+\vec{p}_b^2\bigr\}
 \end{eqnarray}
and 
\begin{equation}
\label{Vi}
	\langle K_s(\vec{k})|V_i|B_s(0)\rangle =H_i=\int d\vec{p}_b\ C(\vec{p}_b)\ \big[E_b(\vec{p}_b)+m_b\big]k_i	\ \ ,
\end{equation}
  
  \noindent where 
  \begin{eqnarray}
  \label{integral}
  	&&C(\vec{p}_b)=\frac{{\cal G}_{B_s}(\vec{p}_b,-\vec{p}_b){\cal G}_{K_s}(\vec{p}_b+\vec{k},-\vec{p}_b)}{\sqrt{N_{B_s}(0)N_{K_s}(\vec{k})}}\nonumber\\
   &&\sqrt{\frac{\big[E_b(\vec{p}_b)+E_s(-\vec{p}_b)\big]\big[E_d(\vec{p}_b+\vec{k})+E_s(-\vec{p}_b)\big]}{E_b(\vec{p}_b)E_d(\vec{p}_b+\vec{k})\big[E_b(\vec{p}_b)+m_b\big]\big[E_d(\vec{p}_b+\vec{k})+m_d\big]}}
  \end{eqnarray}

Now a comparison of the results (\ref{V0},\ref{Vi},\ref{integral}) with the corresponding expression of the covariant factorized amplitude (\ref{Lor_inv_ff_pseudo},\ref{fp},\ref{fm}) yields the Lorentz invariant form factors $f_{\pm}(q^2)$ in the form:
\begin{eqnarray}
\label{F_pm}
	f_{\pm}(q^2)=\frac{1}{2}\int d\vec{p}_b C(\vec{p}_b)&&\biggl\{\big[E_b(\vec{p}_b)+m_b\big]\big[E_d(\vec{p}_b+\vec{k})+m_d\big]+\vec{p}_b^2\nonumber\\
	&&\pm\big[E_b(\vec{p}_b)+m_b\big]\big[M\mp E_{P_1}\big]\biggr\}
\end{eqnarray} 

\subsection{Decay rate of \texorpdfstring{$B_s\to J/\psi K_s$}{Lg}}

In this decay case, the decay amplitude is derivable in terms of invariant form factor $F_1(q^2)$ in the $B_s$-rest frame as
\begin{eqnarray}
	\langle J/\psi K_s|{\cal H}_{eff}|B_s(0)\rangle\;&&= i\frac{G_F}{\sqrt{2}}V_{bc}V_{cd}2a_2\nonumber\\
 &&\times\; m_{J/\psi}f_{J/\psi}F_1(q^2)(e^*.p)
\end{eqnarray}
Here, $e^*$ denotes the polarization vector associated with the daughter meson$(J/\psi)$ and we can get the model expression of $F_1(q^2)$ from eq.(\ref{fp}) and eq.(\ref{F_pm}) as:
\begin{eqnarray}
\label{FF_PV}
	F_1(q^2)&&=f_+(q^2)=\frac{1}{2} \int d\vec{p}_b C(\vec{p}_b)\nonumber\\
 &&\biggl\{\big[E_b(\vec{p}_b)+m_b\big]\big[E_d(\vec{p}_b+\vec{k})+m_d\big]+\vec{p}_b^2+\nonumber\\
 &&\big[E_b(\vec{p}_b)+m_b\big]\big[M-E_P\big]\biggr\}
\end{eqnarray}
 Then the decay rate $\Gamma(B_s\to J/\psi K_s)$ is expressed as 
\begin{equation}
\label{DW_J_psi}
	\Gamma(B_s\to J/\psi  K_s)=\frac{|\vec{k}|^3}{8\pi m_{J/\psi }^2}|A_2|^2|F_1(q^2)|^2,
\end{equation}
where
\begin{equation}
	|A_2|=\frac{G_F}{\sqrt{2}}V_{bc}V_{cd}2a_2m_{J/\psi }f_{J/\psi}
\end{equation}

\subsection{Decay rate of \texorpdfstring{$B_s\to \eta_c K_s$}{Lg}}
For the class-II category of $B_s\to \eta_c K_s$ the decay amplitude in $B_s$ rest frame is written as 
\begin{eqnarray}
    \langle \eta_cK_s|{\cal H}_{eff}|B_s(0)\rangle\;&&= i\frac{G_F}{\sqrt{2}}V_{bc}V_{cd}a_2(M-m_{K_s})\nonumber\\
 &&\times\; f_{\eta_c}F_0(q^2),
\end{eqnarray}
  where
  \begin{equation}
  \label{F0}
  F_0(q^2)=\bigg[\frac{q^2}{(M^2-m_{K_s}^2)}\bigg]f_-(q^2)+f_+(q^2)    
  \end{equation}
  
  From eq.(\ref{F_pm}), it is straightforward to get the model expression in terms of which the decay rate $\Gamma(B_s\to \eta_c K_s)$ is obtained as 
  \begin{equation}
  \label{DW_eta_c}
  	\Gamma(B_s\to \eta_c K_s)=\frac{|\vec{k}|}{8\pi M^2}|A_1|^2|F_0(q^2)|^2,
  \end{equation}
where
\begin{equation}
	|A_1|=\frac{G_F}{\sqrt{2}}V_{bc}V_{\bar{q}_iq_j}a_1(M^2-m_{K_s}^2)f_{\eta_c}
\end{equation}

\section{Numerical results and discussion}
\label{Num_analysis}
To compute the two-body nonleptonic $B_s$ decays within the RIQ model, it is necessary to determine the flavor-independent potential parameters $(a, V_0)$, quark masses $(m_q)$, and their respective binding energies $(E_q)$. Notably, these parameters have been previously established in our model to accurately reproduce experimental meson spectra across both light and heavy flavors sector\cite{Barik:1986mq,Barik:1993aw,Barik:1987zb} and subsequently used in the description of a wide ranging hadronic phenomena\cite{Barik:1992pq,Barik:1994vd,Priyadarsini:2016tiu,Barik:1995sq,Barik:2001gr,Barik:1996kn,Barik:1993yj,Barik:1993aw,Barik:2008zza,Barik:2008zz,Barik:2009zza,Barik:1996xf,Barik:1997qq,Barik:2009zz,Patnaik:2017cbl,Patnaik:2018sym,Patnaik:2019jho,Nayak:2021djn,Barik:2001vp,Barik:2009zzb,Naimuddin:2012dy,Kar:2013fna} involving participating mesons in their ground state. Consequently, the potential parameters employed in this study are:
\begin{equation}
\label{potential_parameter}
(a,V_0)=(0.017166\ GeV^3,-0.1375\ GeV)
\end{equation}

\noindent The quark masses and corresponding binding energies in $GeV$ are taken as: 
\begin{eqnarray}
\label{mass_BE}
m_d=&&0.07875 \ \ \ \ \ \ \ \ \ \ \ \ \ \ \ E_d=0.47125\nonumber\\ 
m_s=&&0.31575 \ \ \ \ \ \ \ \ \ \ \ \ \ \ \ E_s=0.591\nonumber\\
m_c=&&1.49275 \ \ \ \ \ \ \ \ \ \ \ \ \ \ \ E_c=1.57951\nonumber\\
m_b=&&4.77659 \ \ \ \ \ \ \ \ \ \ \ \ \ \ \ E_b=4.76633
\end{eqnarray}

\noindent For relevant CKM parameters and lifetime of $B_s$-meson, we take their values from PDG \cite{ParticleDataGroup:2022pth} as:
\begin{eqnarray}
|V_{cb}|=&&0.0408\pm 0.0014,\ |V_{cd}|=0.221\pm 0.004,\nonumber\\
\tau_{B_s}=&&1.521\pm 0.005\ ps 
\end{eqnarray}
For the masses and decay constants of the involved mesons, which are regarded as phenomenological inputs in this analysis, we adopt their values from the existing observed data \cite{ParticleDataGroup:2022pth} and lattice computation\cite{Becirevic:2016rfu,Dudek:2006ej} respectively. Accordingly, the updated meson masses (MeV) and decay constant(MeV) used in the present analysis are shown in the following Table \ref{mass_decay_constant}.
\begin{table}[hbt!]
    \centering
    \setlength\tabcolsep{1.3pt}
    \renewcommand{\arraystretch}{2.0}
    \begin{tabular}{c|c|c}
    \hline
         Mesons& Mass(in MeV)\cite{ParticleDataGroup:2022pth}&Decay constant (in MeV) \\
         \hline
         \hline
         $B_s$&$5366.92\pm 0.10$&-\\
         $K_s$&$497.61\pm0.013$&-\\
         $J/\psi$ & $3096.9\pm 0.006$&$418\pm8\pm5$\cite{Becirevic:2016rfu}\\
         $\eta_c$&$2983.9\pm 0.04$&$387\pm 7\pm 2$\cite{Becirevic:2016rfu} \\
         $\psi(2S)$&$3686.1\pm0.06$&$143\pm81$\cite{Dudek:2006ej}\\
          $\eta_c(2S)$&$3637.7\pm 1.1$&$56\pm21\pm3$\cite{Dudek:2006ej}\\
        \hline
        \hline
    \end{tabular}
    \caption{Masses and Decay constants of participating mesons.}
    \label{mass_decay_constant}
\end{table}
As mentioned earlier, our calculations utilize the potential parameters (\ref{potential_parameter}) and the quark masses along with their binding energies (\ref{mass_BE}) for ground states. These parameters were previously determined at the static level application of the RIQ model by fitting the mass spectra of ground-state mesons \cite{Barik:1986mq,Barik:1987zb}. The same parameter set has been consistently applied in prior studies employing the RIQ model, providing a satisfactory description of various hadronic phenomena across both light and heavy flavor mesons in their ground state. Hence, our approach does not involve any adjustable parameters that would require fine-tuning for predicting a broad range of hadronic phenomena, as mentioned earlier. Therefore, our studies involve parameter free calculations.\\

Concerning the QCD coefficient $a_2$, various values have been employed in existing literature when computing the nonleptonic transitions of $B_s$ mesons resulting from $b$-quark decay. For example, Colangelo $\it{et al.}$ in \cite{Colangelo:1999zn} use QCD co-efficient $(a_2^b)=(-0.26)$ as fixed in \cite{Browder:1996af} whereas, S. Dubnicka $\it{et al.}$ \cite{Dubnicka:2017job}, use the QCD co-efficient $(a_2^b)=(-0.27)$ for the calculation of BF. We have used the value of $a_2=-0.27$ in our calculation.\\

\begin{figure}[hbt!]
\centering
\includegraphics[width=0.5\textwidth]{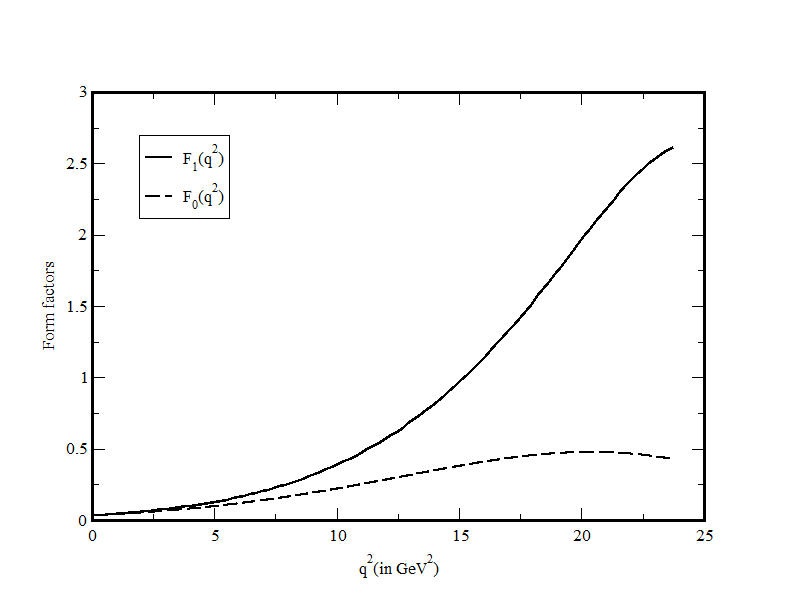}
\caption{$q^2$-dependence of form factors for the decay $B_s \to J/\psi(\eta_c) K_s$.}
\label{ff_graph}
\end{figure}

Utilizing the model parameters provided in equations (\ref{potential_parameter}) and (\ref{mass_BE}), our initial investigation focuses on examining the $q^2$-dependence of the form factors within the permissible kinematic region: $0 < q^2 \leq q^2_{\text{max}}$, as dictated by the analytic expression (\ref{F_pm},\ref{F0},\ref{FF_PV}). Employing a self-consistent dynamic approach, we derive the form factors through the overlapping integrals of meson wave functions, thereby inherently encoding the $q^2$-dependence in the relevant expressions. The $q^2$-dependence of the form factors $F_1(q^2)$ and $F_0(q^2)$ for nonleptonic $B_s$ decays to $K_s$ daughter meson in ground state is illustrated in Figure \ref{ff_graph}. Our analysis reveals that the form factors associated with transitions to $1S$ meson states exhibit a consistent rise with increasing $q^2$ across the entire kinematic spectrum. This observation aligns with the understanding that the $B_s \to J/\psi K_s$ and $B_s\to \eta_c K_s$ transition displays a larger $q^2$-dependence, as the relevant form factors $F_1(q^2)$ and $F_0(q^2)$ exhibit an increase in magnitude with rising $q^2$, encompassing strong $q^2$ dependence. We also observe that the form factor $F_1(q^2)$ is dominated over $F_0(q^2)$ through out the kinematic range for the $B_s\to K_s$ transition. Therefore, it is evident that, the decay rate of $B_s\to \psi(nS) K_s$ is greater than $B_s\to \eta_c(nS) K_s$.

We then compute the BF, for the nonleptonic transitions of $B_s$ to ground as well as radially excited $2S$ charmonium  with a $K_s$ meson, using the expressions provided in eq.(\ref{DW_eta_c},\ref{DW_J_psi}). Our results for decays of $B_s$ meson corresponding to the QCD coefficients $(a_2)$ from the operator product expansion are presented in Table \ref{BF_results}. Our predicted BF for transitions to 2S final meson state are found about one and two order of magnitude smaller than those for 1S final state of vector and pseudoscalar charmonium, respectively. This is due to the fact that, for decays involving higher excited charmonia, the decay constants extracted by \cite{Dudek:2006ej} are smaller than that for ground state \cite{Becirevic:2016rfu}. In our findings, the error ranges are determined by taking upper and lower uncertainties of input parameters, respectively. Compared to the experimental estimation of  LHCb Collaboration for $B_s\to J/\psi K_s$, our result is in good agreement as can be seen in the Table \ref{tab:my_label}. For $B_s\to \psi(2S) K_s$, our prediction is consistent with the CMS data within the experimental uncertainties. As we are resorting to the parameter free calculation from the model framework, therefore the dominant theoretical errors stem solely from the uncertainties of the input parameters $|V_{cb}| \text{ and } |V_{cd}|$, meson masses, lifetime of $B_s$-meson and decay constants of charmonium states.
\begin{table}[hbt!]
    \centering
    \label{BF_results}
    \setlength\tabcolsep{1.3pt}
    \renewcommand{\arraystretch}{2.0}
    \begin{tabular}{|c|c|c|}
         \hline
         Decays&Our work& Exp. results \\
         \hline 
         \hline
         $B_s\to J/\psi K_s$&$2.023_{-0.295}^{+0.325}$ &$1.92 \pm 0.14$\cite{ParticleDataGroup:2022pth}\\
         \hline
         $B_s\to \eta_c K_s$&$1.192_{-0.162}^{+0.184}$&-\\
         \hline
         $B_s\to \psi(2S)K_s$&$0.515_{-0.506}^{+0.89}$&$0.97\pm 0.20(stat)\pm 0.03(syst)$\\
         &&$\pm 0.22(\frac{f_s}{f_d})$\cite{CMS:2022cot}\\
         
         \hline$B_s\to \eta_c(2S)$&$0.0573_{-0.0375}^{+0.0638}$&-\\
         \hline
    \end{tabular}
    \caption{Comparison of BF (in $10^{-5}$) of the $B_s\to \psi (nS) K_s $ and $B_s\to \eta_c(nS) K_s$ decays with experimental results.}
    \label{tab:my_label}
\end{table}
\begin{table}[!hbt]
    \centering
    
    \addtolength{\tabcolsep}{-1pt}
    \begin{tabular}{|c|c|c|}
        \hline
         $\mathcal{R}$&Our Result& Exp. Result  \\
        \hline
        
        \hline &&\\
         $\frac{{\cal B}(B_s\to J/\psi K_s)}{{\cal B}(B_d\to J/\psi K_s)}$&$0.0454^{+0.006}_{-0.0057}$&$.0431\pm 0.0017\; (stat)\pm 0.0012\;(syst)
    \pm 0.0025$\cite{LHCb:2015brj}\\
         \hline
          $\frac{{\cal B}(B_s\to \psi(2S) K_s)}{{\cal B}(B_d\to \psi(2S) K_s)}$&$0.0177^{+0.0269}_{-0.0173}$&$0.033\pm 0.0069\ (stat)\pm 0.0011(syst)$\cite{CMS:2022cot}\\
         &&\\
         \hline
    \end{tabular}
\caption{RIQ model predictions of ratios of branching fractions in comparison with Experimental observations.}
\label{tab:res}
\end{table}

In view of exisiting experimental results, finally, in Table \ref{tab:res}, we report our result for the observable $\mathcal{R}$: 
$\mathcal{B}(B_s \to J/\psi K_s)$/$\mathcal{B}(B_d \to J/\psi K_s)$ along with the 2S state of charmonium as well. It may be noted that CKM matrix $V_{cb}$ and decay constats do not contribute to the ratio ${\cal R}$. The QCD parameter which appears in the decay amplitudes and theoretical uncertainties caused by naive factorization for nonleptonic decay get canceled a lot in calculating the observable $\cal R$. So, the above mentioned ratio provide an essential test of the decays currently under study. By utilizing the world average values of ${\cal B}(B_d\to \psi(nS)K_s)$ \cite{ParticleDataGroup:2022pth}, we calculate the ratio and compared it with experimental findings, revealing a strong concordance between the two. These findings validate our approach and underscore the precision achievable in theoretical predictions for these decays, considering the uncertainties in input parameters.

As we embark on an era of Belle II and the LHCb upgrades \cite{Barel:2022wfr,Fleischer:2022axm}, it becomes evident that numerous intriguing and promising advancements await us, both in experimental and theoretical realms. With the advance of precision measurements these endeavors may serve as a gateway to exploring physics in $B_s$ systems.

\section{Summary and conclusion}
\label{sum_con}
In this work, we investigate the specific cases of exclusive two-body nonleptonic decays, focusing on $B_s\to \psi(1S,2S) K_s$ and $B_s\to \eta_c(1S,2S) K_s$. Specifically, we examine the decay categories $B_s\to PV$ and $B_s\to PP$ within the framework of RIQ model, employing a flavor-independent interaction potential expressed in a scalar-vector harmonic form. Utilizing the factorization approximation, we extract weak decay form factors, representing decay amplitudes and their dependence on the momentum transfer squared $q^2$ across the entire kinematic range $0\leq q^2\leq q^2_{\text{max}}$. These form factors are derived from the overlapping integrals of meson wave functions within the RIQ model. We first provide $q^2$ distribution spectra of the relevant form factors involving in the current analysis. Our investigation indicates that the form factors corresponding to transitions involving $1S$ meson states consistently increases as $q^2$ rises throughout the entire kinematic range. This finding is in line with the notion that the $B_s \to K_s$ transition demonstrates a pronounced $q^2$-dependence, as evidenced by the growing magnitude of the relevant form factors with increasing $q^2$. The dominance of the form factor $F_1(q^2)$ over $F_0(q^2)$ through out the kinematic range, is due to the phase space expectations. Therefore, it is clear that, the decay rate of $B_s\to \psi(nS) K_s$ is higher than $B_s\to \eta_c(nS) K_s$.\\

The resulting BFs of $B_s\to \psi(1S,2S) K_s$ obtained from our work align with experimental results, as reported by the LHCb and CMS Collaboration. On the other hand, our predictions for $B_s\to \eta_c(nS) K_s$ will become a key reference for both experimentalists and theorists, given the current lack of data in this sector. To ensure thoroughness, additionally, we compute the ratio of BF of $B_s$ to $B_d$ for the same final states, $\psi (nS)$ with $K_s$,  utilizing the world average decay rate of $B_d\to \psi (nS) K_s$ and then compared it with the experimental findings. Our predictions in this analysis, achieved through a parameter-free unification within our model framework, aim to strengthen confidence in SM predictions and contribute to valuable discussions on the crucial $B_s$ sector in flavor physics.

\vspace{-5mm}
\appendix
\section{CONSTITUENT QUARK ORBITALS AND MOMENTUM PROBABILITY AMPLITUDES}\label{app}

In RIQ model a meson is picturised as a color-singlet assembly of a quark and an antiquark independently confined by an effective and average flavor independent potential in the form:
$U(r)=\frac{1}{2}(1+\gamma^0)(ar^2+V_0)$, where ($a$, $V_0$) are the potential parameters. It is believed that the zeroth order quark dynamics  generated by the phenomenological confining potential $U(r)$ taken in equally mixed scalar-vector harmonic form can provide adequate tree level description of the decay process being analyzed in this work. With the interaction potential $U(r)$ put into the zeroth order quark Lagrangian density, the ensuing Dirac equation admits static solution of positive and negative energy as: 
\begin{eqnarray}
\psi^{(+)}_{\xi}(\vec r)\;&=&\;\left(
\begin{array}{c}
\frac{ig_{\xi}(r)}{r} \\
\frac{{\vec \sigma}.{\hat r}f_{\xi}(r)}{r}
\end{array}\;\right)U_{\xi}(\hat r)
\nonumber\\
\psi^{(-)}_{\xi}(\vec r)\;&=&\;\left(
\begin{array}{c}
\frac{i({\vec \sigma}.{\hat r})f_{\xi}(r)}{r}\\
\frac{g_{\xi}(r)}{r}
\end{array}\;\right){\tilde U}_{\xi}(\hat r),
\end{eqnarray}
where $\xi=(nlj)$ represents a set of Dirac quantum numbers specifying 
the eigen-modes;
$U_{\xi}(\hat r)$ and ${\tilde U}_{\xi}(\hat r)$
are the spin angular parts given by,
\vspace{-0.75mm}
\begin{eqnarray}
U_{ljm}(\hat r) &=&\sum_{m_l,m_s}<lm_l\;\frac{1}{2}m_s|
jm>Y_l^{m_l}(\hat r)\chi^{m_s}_{\frac{1}{2}}\nonumber\\
{\tilde U}_{ljm}(\hat r)&=&(-1)^{j+m-l}U_{lj-m}(\hat r)
\end{eqnarray}
With the quark binding energy $E_q$ and quark mass $m_q$
written in the form $E_q^{\prime}=(E_q-V_0/2)$,
$m_q^{\prime}=(m_q+V_0/2)$ and $\omega_q=E_q^{\prime}+m_q^{\prime}$, one 
can obtain solutions to the resulting radial equation for 
$g_{\xi}(r)$ and $f_{\xi}(r)$in the form:
\begin{eqnarray}
g_{nl}&=& N_{nl} (\frac{r}{r_{nl}})^{l+l}\exp (-r^2/2r^2_{nl})
L_{n-1}^{l+1/2}(r^2/r^2_{nl})\nonumber\\
f_{nl}&=& N_{nl} (\frac{r}{r_{nl}})^{l}\exp (-r^2/2r^2_{nl})\nonumber\\
&\times &\left[(n+l-\frac{1}{2})L_{n-1}^{l-1/2}(r^2/r^2_{nl})
+nL_n^{l-1/2}(r^2/r^2_{nl})\right ],
\end{eqnarray}
where $r_{nl}= a\omega_{q}^{-1/4}$ is a state independent length parameter, $N_{nl}$
is an overall normalization constant given by
\begin{equation}
N^2_{nl}=\frac{4\Gamma(n)}{\Gamma(n+l+1/2)}\frac{(\omega_{nl}/r_{nl})}
{(3E_q^{\prime}+m_q^{\prime})}
\end{equation}
and
$L_{n-1}^{l+1/2}(r^2/r_{nl}^2)$ etc. are associated Laguerre polynomials. The radial solutions yields an independent quark bound-state condition in the form of a cubic equation:
\begin{equation}
\sqrt{(\omega_q/a)} (E_q^{\prime}-m_q^{\prime})=(4n+2l-1)
\end{equation}
The solution of the cubic equation provides the zeroth order binding energies of 
the confined quark and antiquark for all possible eigenmodes. 
In the relativistic independent particle picture of this model, the constituent quark and antiquark are thought to move independently inside the $B_s$-meson bound state with momentum $\vec p_b$ and $\vec p_s$, respectively. Their individual momentum probability amplitudes are obtained in this model via momentum projection of respective quark orbitals (A1) in following forms.
For ground state mesons:($n=1$,$l=0$)

\begin{eqnarray}
G_b(\vec p_b)&&={\frac{i\pi {\cal N}_b}{2\alpha _b\omega _b}}
\sqrt {\frac{(E_{p_b}+m_b)} {E_{p_b}}}(E_{p_b}+E_b)\nonumber\\
&&\times\exp {(-{\frac{\vec {p_b}^2} {4\alpha_b}})}\nonumber\\
{\tilde G}_s(\vec p_s)&&=-{\frac{i\pi {\cal N}_s}{2\alpha _s\omega _s}}
\sqrt {\frac{(E_{p_s}+m_s)}{E_{p_s}}}(E_{p_s}+E_s)\nonumber\\
&&\times\exp {(-{\frac{{\vec {p_s}}^2}{4\alpha_s}})}
\end{eqnarray}
The binding energies of constituent quark and antiquark for the ground state of $B_s$ meson as well as the final meson states for $n=1$; $l=0$ can also be obtained by solving respective cubic equations representing appropriate bound state conditions.

\begin{acknowledgments}
\vspace{-1mm}
We express our gratitude to Professors, N. Barik, P. C. Dash, and S. Kar for their helpful discussions and substantial contributions to the development of the RIQ model framework. We also acknowledge useful discussions with Prof. Rukmani Mohanta, during the initial stage of this project. We thank the NISER, Department of Atomic Energy, India, for their financial support.
\end{acknowledgments}
\bibliography{bibliography}{}
\bibliographystyle{utphys}

\end{document}